\documentclass[showpacs,prd,aps,10pt,nofootinbib,final]{revtex4-1}
\usepackage{amsmath}
\usepackage{graphicx}
\usepackage{amssymb}
\usepackage{psfrag}

\begin{document}

\title{Eccentric first post-Newtonian waveforms for compact binaries in
frequency domain with Hansen coefficients}
\author{Bal\'{a}zs Mik\'{o}czi$^{1}$, P\'{e}ter Forg\'{a}cs$^{1,2}$ and M\'{a}ty\'{a}s Vas\'{u}th$^{1}$}

\affiliation{$^{1}$ Research Institute for Particle and Nuclear Physics, Wigner RCP
H-1525 Budapest 114, P.O. Box 49, Hungary}
\affiliation{$^2$LMPT, CNRS-UMR 6083, Universit\'e de Tours, Parc de Grandmont, 37200
Tours, France}

\begin{abstract}
The inspiral and merger of supermassive black hole binary systems with high
orbital eccentricity are among the promising sources of the advanced
gravitational wave observatories. In this paper we derive analytic
ready-to-use first post-Newtonian eccentric waveform in Fourier domain with
the use of Hansen coefficients. Introducing generic perturbations of
celestial mechanics we have generalized the Hansen expansion to the first
post-Newtonian order which are then used to express the waveforms. Taking
into account the high eccentricity of the orbit leads to the appearance of
secular terms in the waveform which are eliminated with the introduction of
a phase shift. The waveforms have a systematic structure and as our main
result these are expressed in a tabular form.
\end{abstract}

\maketitle

\section{Introduction}

Compact binaries (i. e. black holes, neutron stars and white dwarfs) with
non-vanishing eccentricity are promising sources of gravitational waves.
Depending on their parameters these sources emit radiation
within the sensitivity band of the forthcoming gravitational wave detectors
advanced LIGO and Virgo. Due to their increased sensitivity the
source signals will be visible by these detectors for a longer time period
with the requirement of an accurate description of both the orbital
evolution and the emitted GWs of these systems.
detectability
The measured signal output of the detectors is cross correlated with
theoretical waveform templates in matched filtering. The presence of
orbital eccentricity changes significantly the properties of the
waveforms resulting in the decrease of their detectability with the
use of circular templates.In spite of the general circularization of
binary orbits due to GW emission binaries interacting with their
environment can retain non-negligible eccentricity at the end of
their evolution. For example, there are indications that binaries in
dense galactic nuclei \cite{LearyKocsis,KocsisLevin}, embedded in a
gaseous disk \cite{Cuadra,Sesana} can remain eccentric \ until the
end of their inspiral. Moreover, the interaction of supermassive
black hole binaries with star populations \cite{Preto,Lockman} and
the Kozai mechanism and relativistic orbital resonances in
hierarchial triples \cite{Wen,Hoffman,Seto,NaozKocsis} can also
increase orbital eccentricity.

The first analytic eccentric waveform for the unperturbed motion was given
by \cite{Whalquist}. The time-dependent waveform is computed with the help
of Fourier-Bessel method in Ref. \cite{Moreno} and the waveform in Fourier
space for arbitrary eccentricity was given in Ref \cite{mikoczi}.

Recently, within the post-Newtonian (PN) treatment of compact binary
evolution the theoretical computations are reaching the level of
4PN. The first post-Newtonian eccentric waveform for bound orbits is
computed by Wagoner and Will \cite{Wagoner} in 1976. The description
of the Kepler motion with the 1PN correction is given by Damour and
Deruelle in Ref \cite{DD} which requires three eccentricities
(\textit{radial},\textit{\ time} and \textit{angular}
\textit{eccentricity}). With the help of this Damour-Deruelle
parameterization the eccentric waveform and evolution of the
\textit{semimajor axis} and the radial eccentricity due to radiation
reaction is computed by Junker and Sch\"{a}fer in Ref \cite{Junker}.
The waveform in Fourier domain up to 1PN and 2PN orders are given in
Refs. \cite {Tessmer},\cite{Tessmer2}.

Our work focuses on ready-to-use eccentric 1PN waveforms in frequency
domain. In our description we use the generalized true anomaly
parameterization which has the advantage that the solution of the equations
of motion can be expressed with only two eccentricities. As a consequence,
the gravitational waveforms have a simple structure. Secular terms appearing
in the waveforms are eliminated by the use of the Poincar\'{e}-Lindstedt
method and the introduction of the \textit{drift true anomaly} parameter. We
give both the time-dependent and frequency domain waveforms using the Hansen
expansion and the stationary phase approximation (SPA). Moreover, we compute
the evolution of the semimajor axis and radial eccentricity up to 1PN
accuracy.

The paper is organized as follows. After a short summary of the
Damour-Deruelle parameterization in Sec. II we introduce the generalized
true anomaly parameterization in Sec. III and the 1PN waveform in Sec. IV.
Sec. V. contains the extension of the Hansen coefficients to 1PN order.
Fourier domain SPA waveforms are given in Sec VI. The radiation reaction
problem and the evolution of the time and phase functions to 1PN order are
given in Sec VII. Some of the technical details, i.e. tensor spherical
harmonics, orbital parameters to 1PN order, Hansen coefficients and waveform
expressions are presented in the Appendices.

\section{Damour-Deruelle parameterization}

In the following, we summarize the first post-Newtonian parameterization of
the orbital motion introduced by Damour and Deruelle \cite{DD} for the
description of compact binaries. The Lagrangian of the system contains the
Newtonian and 1PN corrections,
\begin{eqnarray}
\mathcal{L} &\mathcal{=}&\frac{\mu }{2}\mathbf{v}^{2}+\frac{Gm\mu }{r}
\notag \\
&&+\frac{1}{8c^{2}}\left( 1-3\eta \right) \mu \mathbf{v}^{4}+\!\frac{Gm\mu }{%
2rc^{2}}\!\left[ \!\left( 3\!+\!\eta \right) \mathbf{v}^{2}\!+\!\eta \dot{r}%
^{2}\!-\!\frac{Gm}{r}\!\right] \mathbf{\ ,}
\end{eqnarray}%
where $r$ is the relative distance, $\mathbf{v}$ is the relative velocity
vector, $m=m_{1}+m_{2}$ is the total mass, $\mu =m_{1}m_{2}/m$ is the
reduced mass and $\eta =\mu /m$ is the symmetric mass ratio. The overdot
denotes the derivative with respect to time $t$. The radial and angular
motion can be separated to the linear order, so the Euler-Lagrange equation
are
\begin{align}
\left( \frac{dr}{dt}\right) ^{2}& =A+\frac{2B}{r}+\frac{C}{r^{2}}+\frac{D}{%
r^{3}}\mathbf{\ ,}  \label{ddrad} \\
\frac{d\theta }{dt}& =\frac{H}{r^{2}}+\frac{I}{r^{3}}\mathbf{\ ,}
\label{ddteta}
\end{align}%
where $\theta $ is the azimuthal angle in the orbital plane and the
constants $A,B,C,D,H,I$ depend on the conserved quantities such as energy
and the orbital angular momentum of the perturbed motion, see Appendix B. The
constants $A,B,C,H$ contain Newtonian and 1PN terms while $D$ and $I$ are
purely 1PN corrections. The equations of motion, Eqs. (\ref{ddrad},\ref{ddteta})
can be solved by the \textit{eccentric anomaly} quasi-parameterization $u$,
that is
\begin{equation}
r=a_{r}(1-e_{r}\cos u)\mathbf{\ ,}  \label{damour}
\end{equation}
where the orbital parameters are the semimajor axis $a_{r}$ and the radial
eccentricity $e_{r}$. These orbital parameters are characterized by the
turning points ($r_{\max }$ and $r_{\min }$ in \cite{KMG},\cite{KJ}) of the
radial motion. The Kepler equation and angular evolution can be given as
\begin{eqnarray}
n(t-t_{0}) &=&u-e_{t}\sin u\equiv \mathcal{M}\mathbf{\ ,}  \label{Kepler} \\
\theta -\theta _{0} &=&(1+k)v_{\theta }\mathbf{\ ,}  \label{DD_angle} \\
v_{\theta } &=&2\arctan \sqrt{\frac{1+e_{\theta }}{1-e_{\theta }}}\tan \frac{%
u}{2}\mathbf{,}  \label{intro_v}
\end{eqnarray}
in terms of the orbital elements of the 1PN orbital dynamics such as
the \textit{mean motion} $n$, the \textit{time eccentricity}
$e_{t}$, the \textit{angle eccentricity} $e_{\theta }$, and the
\textit{pericenter drift }$k$ (which is in relationship with the
\textit{pericenter precession} $\left\langle
\dot{\gamma}\right\rangle $ averaged over one radial period, see
\cite{mikoczi}). In the above equations $t_{0}$ and $\theta _{0}$
are integration constants and in our calculations we set $\theta
_{0}=t_{0}=0$. The Damour-Deruelle parameterization contains three
eccentricities, which are different in 1PN order only. In Newtonian
order only the Kepler equation and one eccentricity remain.

We note that the Kepler equation remains the same with the inclusion
of the spin-orbit interaction (see ref. \cite{Wex}) but it contains
higher order contributions such as the spin-spin,
quadrupole-monopole and magnetic dipole-dipole interactions
\cite{KMG} or the second post-Newtonian corrections \cite{SW}. For
this reason we use an other parameterization which has only two
eccentricities and the evolution of the azimuthal angle $\theta $ is
similarly simple as Eq. (\ref{DD_angle}).

\section{Generalized true anomaly parameterization}

We introduce the \textit{generalized true anomaly} parameterization $\phi $
(denoted by $\chi $ in \cite{param},\cite{KMG}) as
\begin{equation}
r=\frac{a_{r}(1-e_{r}^{2})}{1+e_{r}\cos \phi }\mathbf{\ .}  \label{mienk}
\end{equation}
This parameterization has a similar form than the Keplerian
parameterization, where the orbital parameters $a_{r}$ and $e_{r}$
include only the leading order Newtonian terms. Eqs. (\ref{mienk})
and (\ref{damour}) lead to the relation $\phi =v_{r}=2\arctan \left[
\left( 1+e_{r}\right) ^{1/2}\left( 1-e_{r}\right) ^{-1/2}\tan \left(
u/2\right) \right] $, cf. Eq. (\ref{intro_v})\textbf{.}

The time evolution of the generalized true anomaly up to 1PN order can be
expressed as
\begin{equation}
\dot{\phi}=\frac{na_{r}^{2}(1-e_{r}^{2})^{3/2}}{r^{2}(1-e_{t}e_{r}+(e_{r}-e_{t})\cos
\phi )}\mathbf{\ .}  \label{phidot}
\end{equation}
The integration of Eq. (\ref{ddteta}) with the help of (\ref{phidot}) leads
to the relation
\begin{equation}
\theta -\theta _{0}=\left( 1+\kappa _{1}\right) \phi +\kappa _{2}\sin \phi
\mathbf{,}  \label{angleeq}
\end{equation}%
where we have introduced the 1PN order quantities $\kappa
_{1}=3Gm/[a_{r}(1-e_{r}^{2})c^{2}]$ and $\kappa _{2}=G\mu
e_{r}/[2a_{r}(1-e_{r}^{2})c^{2}]$. In this parameterization only
radial and time eccentricities ($e_{r}$, $e_{t}$) appear while angle
eccentricity does not. The relationship between $\phi $ and
$v_{\theta }$ is given by Eqs. (\ref{DD_angle}) and (\ref{angleeq})
up to 1PN order as
\begin{equation}
\phi =v_{\theta }-\frac{G\mu e_{r}}{2a_{r}(1-e_{r}^{2})c^{2}}\sin v_{\theta
}.
\end{equation}

In the following we will compute the full eccentric 1PN waveform with the
use of the generalized true anomaly parameterization $\phi $ without the
appearance of secular terms in the expressions and give the time-domain
waveforms using the generalized \textit{Hansen expansion}. Our aim is to
express the full analytic eccentric frequency-domain waveform up to 1PN order.

\section{Waveform}

The first explicit description of the 1PN waveform for binaries was
given by Wagoner and Will in 1976 \cite{Wagoner}. We rewrite their
expressions with help of Ref. \cite{Thorne} and, as a result, the
radiation field up to 1PN order has the following form
\begin{eqnarray}
h_{ij}^{TT} &=&\varepsilon ^{4}\frac{G}{D_{L}}\Biggl\{\sum_{m=-2}^{2}\overset%
{(2)}{I}_{2m}T_{ij}^{E2,2m}+\epsilon \Biggl[\sum_{m=-2}^{2}{\overset{(2)}{S}}%
_{2m}\,T_{ij}^{B2,2m}  \label{waveform} \\
&&+\sum_{m=-3}^{3}\overset{(3)}{I}_{3m}T_{ij}^{E2,3m}\Biggr]\Biggr\}\,.
\end{eqnarray}%
Here $D_{L}$ is the luminosity distance, $\varepsilon =Gm/c^{2}r$ is
the post-Newtonian parameter, $T_{ij}^{E2,km}$and $T_{ij}^{B2,km}$
are the tensorial \textit{electric} and \textit{magnetic} scalar
harmonics which are given by Eq. (2.30d) in \cite{Thorne}. The
quantities $\overset{(k)}{I}_{km}, $ ${\overset{(k)}{S}}_{km}$ are
the $k$th time derivatives of the mass and current multipole
moments. The explicit form of these multipoles was given by Junker
and Sch\"{a}fer in 1PN order with eccentric anomaly $u$ in refs.
\cite{Junker} and \cite{Tessmer}. Later, the authors of
\cite{Tessmer2} have computed the explicit time-dependent multipoles
$\overset{(k)}{I}_{km}$ and ${\overset{(k)}{S}}_{km}$ up to 2PN.
Their waveforms contain no secular terms because the authors are not
using Eq. (\ref{DD_angle}) but the exponents containing $\theta $
have been expressed as series in $\overset{(k)}{I}_{km}, $
${\overset{(k)}{S}}_{km}$.

It is well-known that some secular terms will appear in the eccentric
waveforms if one expands the harmonic functions of the angle $\theta $ in
terms of the generalized true anomaly parameterization. These secular terms
have to be eliminated in the waveforms which requires the introduction of
the \textit{drift true anomaly }$\phi ^{\prime }=\left( 1+\kappa _{1}\right)
\phi $. Then the harmonic functions of $\theta $ can be described in a
perturbative sense (see \cite{Soffel}) as
\begin{eqnarray}
\cos \theta &\approx &\cos \phi ^{\prime }-\kappa _{2}\sin \phi \sin \phi
^{\prime }, \\
\sin \theta &\approx &\sin \phi ^{\prime }+\kappa _{2}\sin \phi \cos \phi
^{\prime },
\end{eqnarray}
which relations will be used to eliminate secular terms in the eccentric
waveforms.

The waveform up to 1PN order for the polarization states are
\begin{equation}
h_{+,\times }(\phi )=h_{+,\times }^{N}(\phi )+h_{+,\times }^{H}(\phi
)+h_{+,\times }^{PN}(\phi )\mathbf{.}
\end{equation}%
The $h_{+,\times }^{N}(\phi )$ is the Newtonian, $h_{+,\times }^{H}(\phi )$
is the half order PN and $h_{+,\times }^{PN}(\phi )$ is the 1PN waveform
(see Appendix D and E). Our 1PN waveform has the well-known structure with
the generalized true anomaly $\phi $ and drift anomaly $\phi ^{\prime }$, as
\begin{eqnarray}
h_{+,\times }^{PN}(\phi ) &=&\sum_{m=0}^{4}\Biggl[\left( c_{m(c4)}^{+,\times
}\cos m\phi +s_{m(c4)}^{+,\times }\sin m\phi \right) \cos 4\phi ^{\prime }
\notag \\
&&+\left( c_{m(s4)}^{+,\times }\cos m\phi +s_{m(s4)}^{+,\times }\sin m\phi
\right) \sin 4\phi ^{\prime }  \notag \\
&&+\left( c_{m(c2)}^{+,\times }\cos m\phi +s_{m(c2)}^{+,\times }\sin m\phi
\right) \cos 2\phi ^{\prime }  \notag \\
&&+\left( c_{m(s2)}^{+,\times }\cos m\phi +s_{m(s2)}^{+,\times }\sin m\phi
\right) \sin 2\phi ^{\prime }  \notag \\
&&+\left( c_{m}^{+,\times }\cos m\phi +s_{m}^{+,\times }\sin m\phi \right) %
\Biggr]\mathbf{,}  \label{PN}
\end{eqnarray}
where the coefficients $c_{m(c4)}^{+,\times }$, $s_{m(c4)}^{+,\times
}$,$c_{m(c2)}^{+,\times }$, $s_{m(c4)}^{+,\times }$,
$c_{m}^{+,\times }$ and $s_{m}^{+,\times }$ depend on$\ $the radial
eccentricity $e_{r}$, the mass parameters $m$, $\mu $, $\eta $ and
the two polar angles $\Theta $ and $\Phi $ of the line of sight (see
Appendix E). We introduce the quantities
\begin{eqnarray}
\phi _{m}^{\pm 4} &=&m\phi \pm 4\phi ^{\prime }=m_{4\pm }\phi \mathbf{\ ,} \\
\phi _{m}^{\pm 2} &=&m\phi \pm 2\phi ^{\prime }=m_{2\pm }\phi \mathbf{\ ,} \\
\phi _{m} &=&m\phi \mathbf{\ ,}
\end{eqnarray}
with the real numbers $m_{4\pm }=m\pm 4\left( 1+\kappa _{1}\right)
$, $m_{2\pm }=m\pm 2\left( 1+\kappa _{1}\right) $ and coefficients
\begin{eqnarray}
C_{m,+,\times }^{\pm 4} &=&\frac{c_{m(c4)}^{+,\times }\mp
s_{m(s4)}^{+,\times }}{2}\mathbf{\ ,}\mathcal{\qquad }S_{m,+,\times }^{\pm
4}=\frac{s_{m(c4)}^{+,\times }\pm c_{m(s4)}^{+,\times }}{2}\mathbf{\ ,} \\
C_{m,+,\times }^{\pm 2} &=&\frac{c_{m(c2)}^{+,\times }\mp
s_{m(s2)}^{+,\times }}{2}\mathbf{\ ,}\mathcal{\qquad }S_{m,+,\times }^{\pm
2}=\frac{c_{m(c2)}^{+,\times }\pm s_{m(s2)}^{+,\times }}{2}\mathbf{\ .}
\end{eqnarray}
We can use the Hansen expansion for the generalized true anomalies as
\begin{equation}
\cos \lambda \phi =\sum_{k=0}^{\infty }C_{k}^{\lambda }\cos k\mathcal{M},%
\mathcal{\qquad }\sin \lambda \phi =\sum_{k=0}^{\infty }S_{k}^{\lambda }\sin
k\mathcal{M}\mathbf{\ ,}
\end{equation}
with $C_{0}^{\lambda }=X_{0}^{0,\lambda }$, $C_{k}^{\lambda
}=X_{k}^{0,\lambda }+X_{-k}^{0,\lambda }$ and $S_{k}^{\lambda
}=X_{k}^{0,\lambda }-X_{-k}^{0,\lambda }$ where $X_{k}^{n,\lambda }$
are the generalized Hansen coefficients for the real number $\lambda
$ (since $\kappa _{1}$ is not integer) up to 1PN order. We note that
in the Keplerian case (where $\lambda $ is integer and
$e=e_{r}=e_{t}=e_{\theta }$) $\cos \lambda \phi $ and $\sin \lambda
\phi $ can be extended by trigonometric functions of $\cos \phi $ $\
$and $\sin \phi $. Using the Fourier coefficients containing
Bessel-functions (see e.g. \cite{Brumberg})\begin{eqnarray}
\cos \phi &=&-e+\frac{2(1-e^{2})}{e}\sum_{k=1}^{\infty }J_{k}(ke)\cos k%
\mathcal{M},  \label{Kepler1} \\
\sin \phi &=&2\sqrt{1-e^{2}}\sum_{k=1}^{\infty }\frac{J_{k}^{\prime }(ke)}{k}%
\sin k\mathcal{M},  \label{Kepler2}
\end{eqnarray}
where $\prime $ denotes the derivative with respect to the
eccentricity $e$. This 'classical' extension leads to an increasing
order of sums for the increasing value of $\lambda $. Note that Eqs.
(\ref{Kepler1}) and (\ref{Kepler2}) are not valid for the 1PN
motion. Therefore we extend Hansen coefficients up to 1PN order in
the next chapter.

\section{Generalization of Hansen coefficients}

The Hansen coefficients are well-known already since the 19th century in
celestial mechanics (see Appendix C). In our description of the
time-dependent waveforms there appear Hansen coefficients therefore it is
important to extend the Hansen expansion up to 1PN order. The $X_{k}^{n,m}$
Hansen coefficients appear in next series
\begin{equation}
\left( \frac{r}{a}\right) ^{n}\exp (im\phi )=\overset{\infty }{\underset{%
k=-\infty }{\sum }}X_{k}^{n,m}\exp (ik\mathcal{M}).
\end{equation}
The definition of Hansen coefficients are
\begin{equation}
X_{k}^{n,m}=\frac{1}{2\pi }\underset{-\pi }{\overset{\pi }{\int }}\left(
\frac{r}{a}\right) ^{n}\exp (imv-ik\mathcal{M})d\mathcal{M}.
\end{equation}%
In the waveform there appear the harmonic functions of $m\phi $ where$\ m$
is not an integer parameter. So we have to generalize the formula of the
\textit{Keplerian} Hansen coefficients \cite{Jarnagin},
\begin{eqnarray}
X_{k}^{n,m} &=&(1+\beta ^{2})^{-n-1}\overset{\infty }{\underset{s=0}{\sum }}%
\overset{\infty }{\underset{t=0}{\sum }}\binom{n-m+1}{s}  \notag \\
&&\times \binom{n+m+1}{t}\left( -\beta \right) ^{s+t}I_{k-m-s+t}(ke),
\label{jarnagin}
\end{eqnarray}
where $p=k-m-s+t$ is a index notation and $I_{p}(z)$ is the contour integral%
\begin{equation}
I_{p}(z)=\frac{1}{2\pi i}\oint u^{-1-p}\exp \frac{z\left( u-u^{-1}\right) }{2%
}du.
\end{equation}
It is evident if $p$ is\ an integer (\textit{i.e.} $m$ is integer)
then $I_{p}(z)=J_{p}(z)$ where $J_{p}(z)$ is the Bessel function
(\textit{e.g.} for the Newtonian waveform see \cite{mikoczi}). If
$p$ is not an integer then $I_{p}(z)=J_{p}(z)+g_{p}(z)$, where
$g_{p}(z)$ is the correction integral (\cite{WW})
\begin{equation}
g_{p}(z)=-\frac{\sin p\pi }{\pi }\underset{0}{\overset{\infty }{\int }}\exp
(-pu-z\sinh u)du
\end{equation}
for $R(z)>0$.

Due to the different eccentricities we have to generalize Hansen
coefficient in a different way for case of 1PN order. The mean
anomaly (see Eq. (\ref{dM}) in Appendix C) is
\begin{equation}
\frac{d\mathcal{M}}{du}=1-e_{t}\cos u.
\end{equation}
We have introduced the complex quantity $y=\exp iu$, then
\begin{eqnarray}
\frac{r}{a} &=&\left( 1+\beta _{r}^{2}\right) ^{-1}(1-\beta _{r}y)(1-\beta
_{r}y^{-1}), \\
\frac{d\mathcal{M}}{du} &=&1-\frac{e_{t}}{2}\left( y+y^{-1}\right) ,
\end{eqnarray}
where $\beta _{r}=\beta (e_{r})$ (see Appendix C, $\beta
=(1-\sqrt{1-e^{2}})/e$ is the parameter in celestial mechanics).
Then the integrand is
\begin{eqnarray}
\left( X_{k}^{n,m}\right) _{PN} &=&\frac{\left( 1+\beta _{r}^{2}\right) ^{-n}%
}{2\pi }\underset{-\pi }{\overset{\pi }{\int }}y^{m-k}(1-\beta
_{r}y^{-1})^{n+m}(1-\beta _{r}y)^{n-m}  \notag \\
&&\times \left( 1-\frac{e_{t}\left( y+y^{-1}\right) }{2}\right) \exp \frac{%
ke_{t}\left( y-y^{-1}\right) }{2}du  \label{intpn}
\end{eqnarray}%
which can be extended in a form of\ an infinite series of the sum.
Then the generalized Hansen coefficients for 1PN order are
\begin{eqnarray}
\left( X_{k}^{n,m}\right) _{PN} &=&(1+\beta _{r}^{2})^{-n}\overset{\infty }{%
\underset{s=0}{\sum }}\overset{\infty }{\underset{t=0}{\sum }}\binom{n-m}{s}%
\binom{n+m}{t}\left( -\beta _{r}\right) ^{s+t}  \notag \\
&&\times \left[ I_{p}(ke_{t})-\frac{e_{t}}{2}\left(
I_{p-1}(ke_{t})+I_{p+1}(ke_{t})\right) \right] .
\end{eqnarray}
If we use a new index notation $j=t-s$ then generalized Hansen coefficients
for 1PN are
\begin{eqnarray}
\left( X_{k}^{n,m}\right) _{PN} &=&(1+\beta _{r}^{2})^{-n}\overset{s_{1}}{%
\underset{s=0}{\sum }}\overset{j_{1}}{\underset{j=-s}{\sum }}\binom{n-m}{s}%
\binom{n+m}{s+j}\left( -\beta _{r}\right) ^{j}  \notag \\
&&\times \left[ I_{p}(ke_{t})-\frac{e_{t}}{2}\left(
I_{p-1}(ke_{t})+I_{p+1}(ke_{t})\right) \right] .  \label{Jarnagin}
\end{eqnarray}
where $p=k-m+j$. We note that for an integer $m$ the square bracket in
second line of Eq. (\ref{Jarnagin}) can be written as%
\begin{equation}
\lbrack ...]=\left( 1-\frac{p}{k}\right) J_{p}(ke_{t})+\frac{\sin (p\pi )}{%
k\pi }.
\end{equation}
Then the explicit time-dependent waveforms, Eq. (\ref{PN}), are
\begin{equation}
h_{+,\times }^{PN}(t)=\sum_{m=0}^{4}\sum_{k=0}^{\infty }\left( \mathcal{C}%
_{k}^{+,\times ,m}\cos k\mathcal{M}+\mathcal{S}_{k}^{+,\times ,m}\sin k%
\mathcal{M}\right) \mathbf{\ ,}  \label{hhh}
\end{equation}%
where
\begin{eqnarray}
\mathcal{C}_{k}^{+,\times ,m} &=&C_{m,+,\times
}^{-4}C_{k}^{m_{4-}}+C_{m,+,\times }^{+4}C_{k}^{m_{4+}}+C_{m,+,\times
}^{-2}C_{k}^{m_{2-}}  \notag \\
&&+C_{m,+,\times }^{+2}C_{k}^{m_{2+}}+c_{m}^{+,\times }C_{k}^{m}, \\
\mathcal{S}_{k}^{+,\times ,m} &=&S_{m,+,\times
}^{-4}S_{k}^{m_{4-}}+S_{m,+,\times }^{+4}S_{k}^{m_{4+}}+S_{m,+,\times
}^{-2}S_{k}^{m_{2-}}  \notag \\
&&+S_{m,+,\times }^{+2}S_{k}^{m_{2+}}+s_{m}^{+,\times }S_{k}^{m},
\end{eqnarray}
These waveform has significantly simpler structure than the corresponding
expressions in \cite{Tessmer}.

\section{Waveform in Fourier space}

The waveform in Fourier space can be described in the
\textit{stationary phase approximation} of the time-dependent
waveform (see Eqs. (B2) and (B3) in the Appendix B of
\cite{mikoczi}). Taking an arbitrary harmonic function
$\mathcal{A}(t)\cos \Phi (t)$, where $\mathcal{A}(t)$, $\Phi (t)$ is
the time-dependent amplitude and phase, respectively, and the
conditions $\mathcal{\dot{A}}/\mathcal{A}\ll \dot{\Phi}$ and
$\ddot{\Phi}\ll \dot{\Phi} ^{2}$ are satisfied, then the Fourier
transform of the function $\mathcal{A}(t)\cos \Phi (t)$ can be
written as
\begin{eqnarray}
\mathcal{F}\left[ \mathcal{A}(t)\sin \Phi (t)\right] &=&\tfrac{\mathcal{A}%
\left[ f(\mathcal{T})\right] }{2}\sqrt{\tfrac{2\pi }{\left\vert \ddot{\Psi}%
\left[ f(\mathcal{T})\right] \right\vert }}e^{i\left( \Psi \left[ f(\mathcal{%
T})\right] +\frac{\pi }{4}\right) }\ ,  \label{int1} \\
\mathcal{F}\left[ \mathcal{A}(t)\cos \Phi (t)\right] &=&\tfrac{\mathcal{A}%
\left[ f(\mathcal{T})\right] }{2}\sqrt{\tfrac{2\pi }{\left\vert \ddot{\Psi}%
\left[ f(\mathcal{T})\right] \right\vert }}e^{i\left( \Psi \left[ f(\mathcal{%
T})\right] -\frac{\pi }{4}\right) }\ ,  \label{int2}
\end{eqnarray}%
where $\Psi \left[ f(\mathcal{T})\right] =2\pi f(\mathcal{T})t\left[
\nu (\mathcal{T})\right] -\Phi \left[ \nu (\mathcal{T})\right] $ is
the phasing function, $\mathcal{T}$ is the saddle point and the
functions $t\left[ \nu (\mathcal{T})\right] $ and $\Phi \left[ \nu
(\mathcal{T})\right] $ appearing in the above expressions can be
obtained from the leading order equations for gravitational
radiation by Appell functions (see the Appendix in \cite{mikoczi}).
It is necessary to add, that here the phase and frequency ($\Psi
_{n}$ and $\Psi _{n\pm }$) are not splitting into triplet due to
pericenter precession (it was a consequence of the appearance of
$\dot{\gamma}$, \textit{i.e.} \textit{heuristic precession} in
\cite{mikoczi}), because it is contained directly the 1PN equations
of motion (see the orbital parameter $k$). Therefore the 1PN
waveform depends on the single phase and frequency $\Psi _{n}$.
Accordingly, the waveform, Eq. (\ref{hhh}), in the Fourier space
becomes
\begin{eqnarray}
h_{+}^{PN}(f) &=&\frac{1}{2}\sqrt{\tfrac{2\pi }{\left\vert \ddot{\Psi}\left[
f(\mathcal{T})\right] \right\vert }}\sum_{m=0}^{4}\sum_{k=0}^{\infty }\Biggl[%
\mathcal{C}_{k}^{+,m}\left[ f(\mathcal{T})\right] e^{i\left( \Psi \left[ f(%
\mathcal{T})\right] -\frac{\pi }{4}\right) }  \notag \\
&&+\mathcal{S}_{k}^{+,m}\left[ f(\mathcal{T})\right] e^{i\left( \Psi \left[
f(\mathcal{T})\right] +\frac{\pi }{4}\right) }\Biggr], \\
h_{\times }^{PN}(f) &=&\frac{1}{2}\sqrt{\tfrac{2\pi }{\left\vert \ddot{\Psi}%
\left[ f(\mathcal{T})\right] \right\vert }}\sum_{m=0}^{4}\sum_{k=0}^{\infty }%
\Biggl[\mathcal{C}_{k}^{\times ,m}\left[ f(\mathcal{T})\right] e^{i\left(
\Psi \left[ f(\mathcal{T})\right] -\frac{\pi }{4}\right) }  \notag \\
&&+\mathcal{S}_{k}^{\times ,m}\left[ f(\mathcal{T})\right] e^{i\left( \Psi %
\left[ f(\mathcal{T})\right] +\frac{\pi }{4}\right) }\Biggr],
\end{eqnarray}
with the stationary phase condition $\left\vert \ddot{\Psi}\left[
f(\mathcal{T})\right] \right\vert =2\pi n\dot{\nu}$ and the phasing
function $\Psi\left[ f(\mathcal{T})\right] =2\pi
ft(\mathcal{T})-\Phi _{n}(\mathcal{T})$. The above form can be
written as
\begin{eqnarray}
h_{+}^{PN}(f) &=&\frac{\left( n\dot{\nu}\right) ^{-1/2}}{2}%
\sum_{m,k=0}^{4,\infty }\Biggl[\mathcal{C}_{k}^{+,m}e^{i\Psi -}+\mathcal{S}%
_{k}^{+,m}e^{i\Psi +}\Biggr],  \label{intt1} \\
h_{\times }^{PN}(f) &=&\frac{\left( n\dot{\nu}\right) ^{-1/2}}{2}%
\sum_{m,k=0}^{4,\infty }\Biggl[\mathcal{C}_{k}^{\times ,m}e^{i\Psi -}+%
\mathcal{S}_{k}^{\times ,m}e^{i\Psi +}\Biggr],  \label{intt2}
\end{eqnarray}
where the phasing functions are $\Psi _{\pm }=2\pi
ft(\mathcal{T})-\Phi _{n}(\mathcal{T})\pm \pi /4$.

Afterwards we shall compute the phase $\Phi _{n}(\mathcal{T})$ and
time $t(\mathcal{T})$ functions appearing in the 1PN waveform.

\section{Radiation reaction to 1PN order}

To leading order the averaged radiative change of the semimajor axis and
eccentricity is governed by the quadrupole formula, see Peters \cite{Peters}.
In these equations the semimajor axis can be replace by the orbital
frequency using Kepler's third law to have the following expressions
\begin{eqnarray}
\dot{\nu}_{N} &=&\frac{48(G\mathcal{M}_{c})^{5/3}(2\pi \nu )^{11/3}}{%
5c^{5}\pi (1-e^{2})^{7/2}}\left( 1+\frac{73}{24}e^{2}+\frac{37}{96}%
e^{4}\right) , \\
\dot{e}_{N} &=&-\frac{304(G\mathcal{M}_{c})^{5/3}(2\pi \nu )^{8/3}}{%
15c^{5}(1-e^{2})^{5/2}}e\left( 1+\frac{121}{304}e^{2}\right) ,
\end{eqnarray}%
where $\mathcal{M}_{c}=m\eta ^{3/5}$ is the chirp mass of the binary system.
The above equations can be integrated and with the use of the exact solution
the phase and time functions can be expressed in terms of the \textit{Appell
functions} (see \cite{mikoczi}). Afterwards, we will compute 1PN corrections
to these equations.

The averaged losses of the radial orbital parameters due the
gravitational radiation reaction up to 1PN order is given by Junker
and Sch\"{a}fer \cite{Junker}. We have to use Kepler's third law in
1PN order relating the orbital frequency and semimajor axis as
\begin{equation}
a_{r}=\frac{\left( Gm\right) ^{1/3}}{(2\pi \nu )^{2/3}}\left( 1+\left( \eta
-9\right) \frac{\left( 2\pi Gm\nu \right) ^{2/3}}{3c^{2}}\right) .
\label{aa}
\end{equation}
The radiative evolution of the orbital frequency and eccentricity up to 1PN
order can be written as (in this chapter we omit the subscript $r$ of the radial eccentricity)
\begin{eqnarray}
\dot{\nu} &=&\dot{\nu}_{N}+\dot{\nu}_{PN},  \label{nudot} \\
\dot{e} &=&\dot{e}_{N}+\dot{e}_{PN},  \label{edot}
\end{eqnarray}
with
\begin{eqnarray}
\dot{\nu}_{PN} &=&\frac{(G\mathcal{M}_{c})^{7/3}\eta ^{-2/5}(2\pi \nu
)^{13/3}}{560c^{7}\pi (1-e^{2})^{9/2}}\Biggl[20368-14784\eta  \notag \\
&&-24e^{2}(2561+2254\eta )-42e^{4}(3885+158\eta )  \notag \\
&&+e^{6}(-13147+1036\eta )\Biggr], \\
\dot{e}_{PN} &=&-\frac{(G\mathcal{M}_{c})^{7/3}\eta ^{-2/5}(2\pi \nu )^{10/3}%
}{2520c^{7}(1-e^{2})^{7/2}}e\Biggl[211944  \notag \\
&&-180320\eta -60e^{2}(11598+1001\eta )  \notag \\
&&+e^{4}(-168303+16940\eta )\Biggr].
\end{eqnarray}%
Thereafter we will find the perturbative solution to the above equations up
to 1PN order.

We get the relation between $\nu $ and $e$ from Eqs. (\ref{nudot})
and (\ref{edot}) up to 1PN order as
\begin{equation}
\frac{d\nu }{de}=\frac{\dot{\nu}_{N}}{\dot{e}_{N}}+\frac{\dot{\nu}_{PN}}{%
\dot{e}_{N}}-\frac{\dot{\nu}_{N}\dot{e}_{PN}}{\dot{e}_{N}^{2}}.
\label{dnude}
\end{equation}%
The exact general solution in the Newtonian order (without the two last
terms in the right hand side of Eq. (\ref{dnude})) is
\begin{equation}
\nu _{N}=\frac{Ce^{-18/19}\left( 1-e^{2}\right) ^{3/2}}{(1+\frac{121}{304}%
e^{2})^{1305/2299}},
\end{equation}%
where $\nu _{N}$ is the general Newtonian solution. Hereafter we use the
expression $\nu _{N}=\nu _{0}\sigma (e)/\sigma (e_{0})$ where the quantities
$\nu _{0}$ and$\ e_{0}$ are initial values for $\nu _{N}(e_{0})=\nu _{0}$
and $\sigma (e)=e^{-18/19}\left( 1-e^{2}\right) ^{3/2}(1+\frac{121}{304}%
e^{2})^{-1305/2299}$ is a shorthand notation. The perturbative Eq.
(\ref{dnude}) has the exact general solution (including the
Newtonian and 1PN
terms) in a mathematical sense%
\begin{equation}
\nu =\left( b_{N}+b_{PN}\right) ^{-3/2},  \label{freqexact}
\end{equation}%
where the quantities $b_{N}$ and $b_{PN}$ are%
\begin{eqnarray}
b_{N} &=&\frac{Ce^{12/19}(1+\frac{121}{304}e^{2})^{870/2299}}{1-e^{2}}, \\
b_{PN} &=&\frac{(2\pi G\mathcal{M}_{c})^{2/3}}{c^{2}\eta ^{2/5}(1-e^{2})}%
\left( 1+\frac{121}{304}e^{2}\right) ^{870/2299}  \notag \\
&&\times \Biggl[\frac{B_{1}+B_{2}e^{2}+B_{3}e^{4}}{(1+\frac{121}{304}%
e^{2})^{3169/2299}}  \notag \\
&&+\,_{2}F_{1}\left( \frac{870}{2299},\frac{13}{19},\frac{32}{19};-\frac{121%
}{304}e^{2}\right) B_{4}e^{2}\Biggr],
\end{eqnarray}%
with the coefficients%
\begin{eqnarray}
B_{1} &=&\frac{1153}{3192}-\frac{89}{114}\eta ,  \notag \\
B_{2} &=&-\frac{2293125927}{558758080}+\frac{60619}{6984476}\eta ,  \notag \\
B_{3} &=&-\frac{86928802699}{93871357440}+\frac{129501097}{670509696}\eta ,
\notag \\
B_{4} &=&\frac{703\,785\,517}{4014\,235\,680}-\frac{49\,913}{735\,208}\eta .
\end{eqnarray}%
Here $_{2}F_{1}\left( \alpha ,\beta ,\gamma ;z\right) $ is the \textit{%
ordinary hypergeometric function}. These general solutions for $\nu (e)$ and
$a(e)$ are consistent with the 1PN order Kepler equation of (\ref{aa}),
strictly speaking if we solve similarly method the original radiation
reaction (equations of $\{\dot{a},$ $\dot{e}\}$ depending on the semimajor
axis) equation up to 1PN order with the use of Eq. (\ref{aa}) then we will
get an identity for semimajor axis $a$ (note that $a=b_{N}$ for Newtonian
order).

Let us identify the Newtonian expression $b_{N}^{-3/2}\equiv \nu
_{N}=C_{0}\sigma (e)$, where $C_{0}=\nu _{0}/\sigma (e_{0})$. The
integration constant $C$ has leading order corrections at 1PN order
so if we require the equation $\nu (e_{0})=\nu _{0}$ to hold, we get
the valid perturbative solution for the orbital frequency, Eq.
(\ref{freqexact}), to 1PN accuracy $\ $
\begin{equation}
\nu =\nu _{0}\frac{\sigma (e)}{\sigma (e_{0})}\left[ 1+\frac{3}{2}\nu
_{0}^{2/3}\left( b_{PN}(e_{0})-\left( \frac{\sigma (e)}{\sigma (e_{0})}%
\right) ^{2/3}b_{PN}(e)\right) \right] .  \label{solfreq}
\end{equation}
Then our aim is to compute the time and phase functions%
\begin{eqnarray}
t-t_{c} &=&\int_{0}^{e}\frac{de^{\prime }}{\dot{e}(e^{\prime })}, \\
\Phi -\Phi _{c} &=&2\pi \int_{0}^{e}\frac{\nu (e^{\prime })}{\dot{e}%
(e^{\prime })}de^{\prime }
\end{eqnarray}
up to 1PN order. The integrals in the Newtonian case is given in
Appendix A of \cite{mikoczi},
\begin{eqnarray}
\left( t-t_{c}\right) _{N} &=&-\frac{15c^{5}\Lambda _{0}^{8/3}}{304(G%
\mathcal{M}_{c})^{5/3}}\left( \frac{\sigma (e_{0})}{2\pi \nu _{0}}\right)
^{8/3}  \notag \\
&&\times \int_{0}^{e}\frac{e^{\prime 29/19}(1-e^{\prime 2})^{-3/2}de^{\prime
}}{\left( 1+\frac{121}{304}e^{\prime 2}\right) ^{-1181/2299}}  \label{tN} \\
\left( \Phi -\Phi _{c}\right) _{N} &=&-\frac{15c^{5}\Lambda _{0}^{5/3}}{304(G%
\mathcal{M}_{c})^{5/3}}  \notag \\
&&\times \int_{0}^{e}\frac{e^{\prime 11/19}}{(1+\frac{121}{304}e^{\prime
2})^{124/2299}}de^{\prime },  \label{PhiN}
\end{eqnarray}
where we have introduced the notation $\Lambda _{0}=\sigma
(e_{0})/(2\pi \nu _{0})$ which depends on the initial eccentricity
and orbital frequency. Such type of integrals can be given by
extended hypergeometric functions (\textit{i.e.} Appell functions),
\begin{eqnarray}
\left( t-t_{c}\right) _{N} &=&-\frac{15c^{5}\Lambda _{0}^{8/3}}{304(G%
\mathcal{M}_{c})^{5/3}}  \notag \\
&&\times \frac{e^{1-\alpha }}{1-\alpha }F_{1}\left( \frac{1-\alpha }{2},\hat{%
\beta},-\gamma ,\frac{3-\alpha }{2};\delta e^{2},e^{2}\right) . \\
\left( \Phi -\Phi _{c}\right) _{N} &=&-\frac{15c^{5}\Lambda _{0}^{5/3}}{304(G%
\mathcal{M}_{c})^{5/3}}  \notag \\
&&\times \frac{e^{1-\tilde{\alpha}}}{1-\tilde{\alpha}}F_{1}\left( \frac{1-%
\tilde{\alpha}}{2},\breve{\beta},0,\frac{3-\tilde{\alpha}}{2};\delta
e^{2},e^{2}\right) ,
\end{eqnarray}%
where $F_{1}\left( \alpha ,\beta ,\beta ^{\prime },\gamma
;x,y\right) $ is the Appell function (see \cite{WW}) and the
constants are $\alpha =-10/19, \tilde{\alpha}=8/19$,
$\breve{\beta}=124/2299,\hat{\beta}=-1181/2299,\gamma
=3/2,$ $\delta =$ $-121/304$ \footnote{%
We have used the following integral formula for the Appell function
$\int_{0}^{x}\frac{(1-x^{\prime 2})^{\gamma }}{x^{\prime \alpha
}\left(
1-\delta x^{\prime 2}\right) ^{\beta }}dx^{\prime }=\frac{x^{1-\alpha }}{%
1-\alpha }F_{1}\left( \frac{1-\alpha }{2},\beta ,-\gamma ,\frac{3-\alpha }{2}%
;\delta x^{2},x^{2}\right) $}. Similiar integrands appear in 1PN order. Then
we can compute the integrand of time function to 1PN order as%
\begin{eqnarray}
\left( t-t_{c}\right) _{PN} &=&-\frac{5c^{3}\Lambda _{0}^{2}}{76G\mathcal{M}%
_{c}\eta ^{2/5}}\int_{0}^{e}\frac{e^{\prime 17/19}(1-e^{\prime 2})^{-3/2}}{%
\left( 1+\frac{121}{304}e^{\prime 2}\right) ^{\allowbreak -1181/2299}}
\notag \\
&&\times \Biggl[\newline
\frac{\tilde{B}_{1}+\tilde{B}_{2}e^{\prime 2}+\tilde{B}_{3}e^{\prime 2}}{%
\left( 1+\frac{121}{304}e^{\prime 2}\right) ^{3169/2299}}  \notag \\
&&+\,_{2}F_{1}\left( \frac{870}{2299},\frac{13}{19},\frac{32}{19};-\frac{121%
}{304}e^{\prime 2}\right) \tilde{B}_{4}e^{\prime 2}\Biggr]de^{\prime }
\notag \\
&&-4\nu _{0}^{2/3}b_{PN}(e_{0})\left( t-t_{c}\right) _{N},  \label{timeint}
\end{eqnarray}
with
\begin{eqnarray}
\tilde{B}_{1} &=&\frac{204288B_{1}-211944+180320\eta }{68096},  \notag \\
\tilde{B}_{2} &=&\frac{204288B_{2}+60(11598+1001\eta )}{68096},  \notag \\
\tilde{B}_{3} &=&\frac{204288B_{3}+168303-16940\eta }{68096},  \notag \\
\,\tilde{B}_{4} &=&3B_{4}.
\end{eqnarray}
Note that in Eq. (\ref{timeint}) the last term is coming from the solution
of the orbital frequency in Eq. (\ref{solfreq}). Computation of the integral
Eq. (\ref{timeint}) is difficult thus we use the approximation $_{2}F_{1}\left( \frac{870}{2299},\frac{13}{19},\frac{32}{19};-\frac{121}{304}%
e^{2}\right) \simeq 1$ because its limit is $1$ for $e\rightarrow 1$ and $%
0.9474$ for $e\rightarrow 1$. Then the time function is
\begin{eqnarray}
\left( t-t_{c}\right) _{PN} &\simeq &-\frac{5c^{3}\Lambda _{0}^{2}}{76G%
\mathcal{M}_{c}\eta ^{2/5}}  \notag \\
&&\times \Biggl[\int_{0}^{e}\frac{e^{\prime 17/19}(1-e^{\prime 2})^{-3/2}(%
\tilde{B}_{1}+\tilde{B}_{2}e^{\prime 2}+\tilde{B}_{3}e^{\prime 4})}{\left( 1+%
\frac{121}{304}e^{\prime 2}\right) ^{\allowbreak 1988/2299}}\newline
de\prime  \notag \\
&&+\tilde{B}_{4}\int_{0}^{e}\,\frac{e^{\prime 55/19}(1-e^{\prime 2})^{-3/2}}{%
\left( 1+\frac{121}{304}e^{\prime 2}\right) ^{\allowbreak -1181/2299}}%
de^{\prime }\Biggr]  \notag \\
&&-4\nu _{0}^{2/3}b_{PN}(e_{0})\left( t-t_{c}\right) _{N}.
\end{eqnarray}
The final result is
\begin{eqnarray}
\left( t-t_{c}\right) _{PN} &\simeq &-\frac{5c^{3}\Lambda _{0}^{2}}{76G%
\mathcal{M}_{c}\eta ^{2/5}}  \notag \\
&&\times \Biggl[\overset{2}{\underset{i=0}{\sum }}\frac{e^{\alpha _{i}}}{%
\alpha _{i}}F_{1}\left( \frac{\alpha _{i}}{2},\beta ,\gamma ,\frac{2+\alpha
_{i}}{2};\delta e^{2},e^{2}\right) \tilde{B}_{i}  \notag \\
&&+\frac{e^{\alpha _{2}}}{\alpha _{2}}F_{1}\left( \frac{\alpha _{2}}{2},\hat{%
\beta},\gamma ,\frac{2+\alpha _{2}}{2};\delta e^{2},e^{2}\right) \tilde{B}%
_{4}\Biggr]  \notag \\
&&-4\Gamma _{0}\left( t-t_{c}\right) _{N},
\end{eqnarray}
where we have introduced the shorthand notations $\gamma =3/2,\delta =$ $%
-121/304,$ $\beta =1988/2299$, $\hat{\beta}=-1181/2299,$ $\ \alpha
_{i}=1-(\alpha _{0}+2i)$, $\alpha _{0}=17/19$ and $\Gamma _{0}=\nu
_{0}^{2/3}b_{PN}(e_{0})$. The phase function $\Phi -\Phi _{c}$ can be
computed similarly. The integrand of the phase function $\Phi -\Phi _{c}$ up
to linear order is \
\begin{eqnarray}
\left( \Phi -\Phi _{c}\right) _{PN} &=&-\frac{5c^{3}\Lambda _{0}}{76G%
\mathcal{M}_{c}\eta ^{2/5}}\int_{0}^{e}\frac{e^{\prime -1/19}}{(1+\frac{121}{%
304}e^{\prime 2})^{124/2299}}  \notag \\
&&\times \Biggl[\frac{\hat{B}_{1}+\hat{B}_{2}e^{\prime 2}+\hat{B}%
_{3}e^{\prime 4}}{(1+\frac{121}{304}e^{\prime 2})^{3169/2299}}  \notag \\
&&+\,_{2}F_{1}\left( \frac{870}{2299},\frac{13}{19},\frac{32}{19};-\frac{121%
}{304}e^{\prime 2}\right) \hat{B}_{4}e^{\prime 2}\Biggr]de^{\prime }  \notag
\\
&&-\frac{15}{2}\Gamma _{0}\left( \Phi -\Phi _{c}\right) _{N},
\label{phaseint}
\end{eqnarray}
where we have introduced the quantities $\hat{B}_{i}=\tilde{B}_{i}-9B_{i}/8$.
We use the above approximation $_{2}F_{1}(..)\simeq 1$, thus the final
form of the phase function is
\begin{eqnarray}
\left( \Phi -\Phi _{c}\right) _{PN} &\simeq &-\frac{5c^{3}\Lambda _{0}}{76G%
\mathcal{M}_{c}\eta ^{2/5}}  \notag \\
&&\Biggl[\overset{2}{\underset{i=0}{\sum }}\frac{e^{\tilde{\alpha}_{i}}}{%
\tilde{\alpha}_{i}}F_{1}\left( \frac{\tilde{\alpha}_{i}}{2},\tilde{\beta},0,%
\frac{2+\tilde{\alpha}_{i}}{2};\delta e^{2},e^{2}\right) \hat{B}_{i}  \notag
\\
&&+\frac{e^{\tilde{\alpha}_{2}}}{\tilde{\alpha}_{2}}F_{1}\left( \frac{\tilde{%
\alpha}_{2}}{2},\hat{\beta},0,\frac{2+\tilde{\alpha}_{2}}{2};\delta
e^{2},e^{2}\right) \hat{B}_{4}\Biggr]  \notag \\
&&-\frac{15\Gamma _{0}}{2}\left( \Phi -\Phi _{c}\right) _{N},
\end{eqnarray}
where $\gamma =3/2,\delta =$ $-121/304,$ $\tilde{\beta}=3293/2299$, $\hat{%
\beta}=-1181/2299,$ $\tilde{\alpha}_{i}=1-(\tilde{\alpha}_{0}+2i)$ and $%
\tilde{\alpha}_{0}=-1/19$. These constants are summarized in Table
I. In summary, the time and phase function up to 1PN are
\begin{eqnarray}
t-t_{c} &=&t_{N}+t_{PN} \\
\Phi -\Phi _{c} &=&\Phi _{N}+\Phi _{PN}
\end{eqnarray}
\begin{eqnarray}
t_{N} &=&-\frac{15c^{5}\Lambda _{0}^{8/3}}{304(G\mathcal{M}_{c})^{5/3}}%
F(e,\alpha ,\hat{\beta},\gamma ,\delta ),  \notag \\
t_{PN} &=&-\frac{5c^{3}\Lambda _{0}^{2}}{76G\mathcal{M}_{c}\eta ^{2/5}}%
\Biggl[\overset{3}{\underset{i=1}{\sum }}F(e,\alpha _{i},\beta ,\gamma
,\delta )\tilde{B}_{i}  \notag \\
&&+F(e,\alpha _{2},\hat{\beta},\gamma ,\delta )\tilde{B}_{4}\Biggr]-4\Gamma
_{0}t_{N},  \notag \\
\Phi _{N} &=&-\frac{15c^{5}\Lambda _{0}^{5/3}}{304(G\mathcal{M}_{c})^{5/3}}%
F(e,\tilde{\alpha},\breve{\beta},0,\delta ),  \notag \\
\Phi _{PN} &=&-\frac{5c^{3}\Lambda _{0}}{76G\mathcal{M}_{c}\eta ^{2/5}}%
\Biggl[\overset{3}{\underset{i=1}{\sum }}F(e,\tilde{\alpha}_{i},\hat{\beta}%
,0,\delta )\hat{B}_{i}  \notag \\
&&+F(e,\tilde{\alpha}_{2},\tilde{\beta},0,\delta )\hat{B}_{4}\Biggr]-\frac{15%
}{2}\Gamma _{0}\Phi _{N}.
\end{eqnarray}%
with $\Lambda _{0}=\sigma (e_{0})/(2\pi \nu _{0})$, $\Gamma _{0}=\nu
_{0}^{2/3}b_{PN}(e_{0})$ and the function $F(e,\alpha ,\beta ,\gamma ,\delta )\doteq
F_{1}\left( \frac{\alpha }{2},\beta ,\gamma ,\frac{2+\alpha }{2};\delta
e^{2},e^{2}\right) e^{\alpha }/\alpha $. \footnote{%
It can be noticed that $F(e,\tilde{\alpha}_{N},\tilde{\beta}_{N},0,\delta
)=_{2}F_{1}(...)$ for $\Phi _{N}$ and $\Phi _{PN}$}

The qualitative behavior of the orbital evolution is presented in
Figures 1-3.

\begin{table}[h]
\caption{Constants of the time and phase functions.}
\label{constants}%
\begin{tabular}{c|c|c}
\hline\hline
$\delta =$ $-\frac{121}{304}$ & N & PN \\ \hline
$t-t_{c}$ & \multicolumn{1}{|l}{$\gamma =\frac{3}{2}$} & \multicolumn{1}{|l}{$%
\gamma =\frac{3}{2}$} \\
& \multicolumn{1}{|l}{$\hat{\beta}=-\frac{1181}{2299}$} & \multicolumn{1}{|l}{%
$\beta =\frac{1988}{2299}$} \\
& \multicolumn{1}{|l}{$\alpha =-\frac{10}{19}$} & \multicolumn{1}{|l}{$\alpha
_{0}=\frac{17}{19}$, $\alpha _{1}=\frac{55}{19}$, $\alpha _{2}=\frac{93}{19}$%
} \\ \hline
$\Phi -\Phi _{c}$ & \multicolumn{1}{|l}{$\gamma =0$} & \multicolumn{1}{|l}{$%
\gamma =0$} \\
& \multicolumn{1}{|l}{$\breve{\beta}=\frac{124}{2299}$} & \multicolumn{1}{|l}{%
$\tilde{\beta}=\frac{3293}{2299}$} \\
& \multicolumn{1}{|l}{$\tilde{\alpha}=\frac{8}{19}$} & \multicolumn{1}{|l}{$%
\tilde{\alpha}_{0}=\frac{1}{19}$, $\tilde{\alpha}_{1}=\frac{37}{19}$, $%
\tilde{\alpha}_{2}=\frac{75}{19}$} \\ \hline\hline
\end{tabular}%
\end{table}

\begin{figure}[h]
\includegraphics[height=4cm]{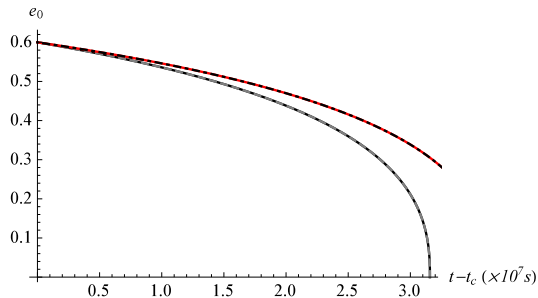}
\caption{(color online). Comparison of the analytical and numerical results for the evolution of the eccentricity. The initial eccentricity is $e_{0}=0.6$, the masses of binary are $m_{i}=10^{6}(1+z)M_{\odot }$ with
redshift $z=1$. The inspiral starts one year before the
\textit{last stable orbit (LSO)} calculated in the Newtonian order which corresponds to the
initial frequency $\protect\nu _{0}=8.09\protect\mu \mathrm{Hz}$. In the Newtonian case the
dotted black line denotes the analytic, while the gray line the numeric solution.
For the 1PN orbital evolution the analytic and numeric solutions are denoted by the dotdashed and red lines, respectively. It can be seen that the perturbative solution is in perfect agreement with the numerical one.}
\label{fig2}
\end{figure}

\begin{figure}[h]
\includegraphics[height=4cm]{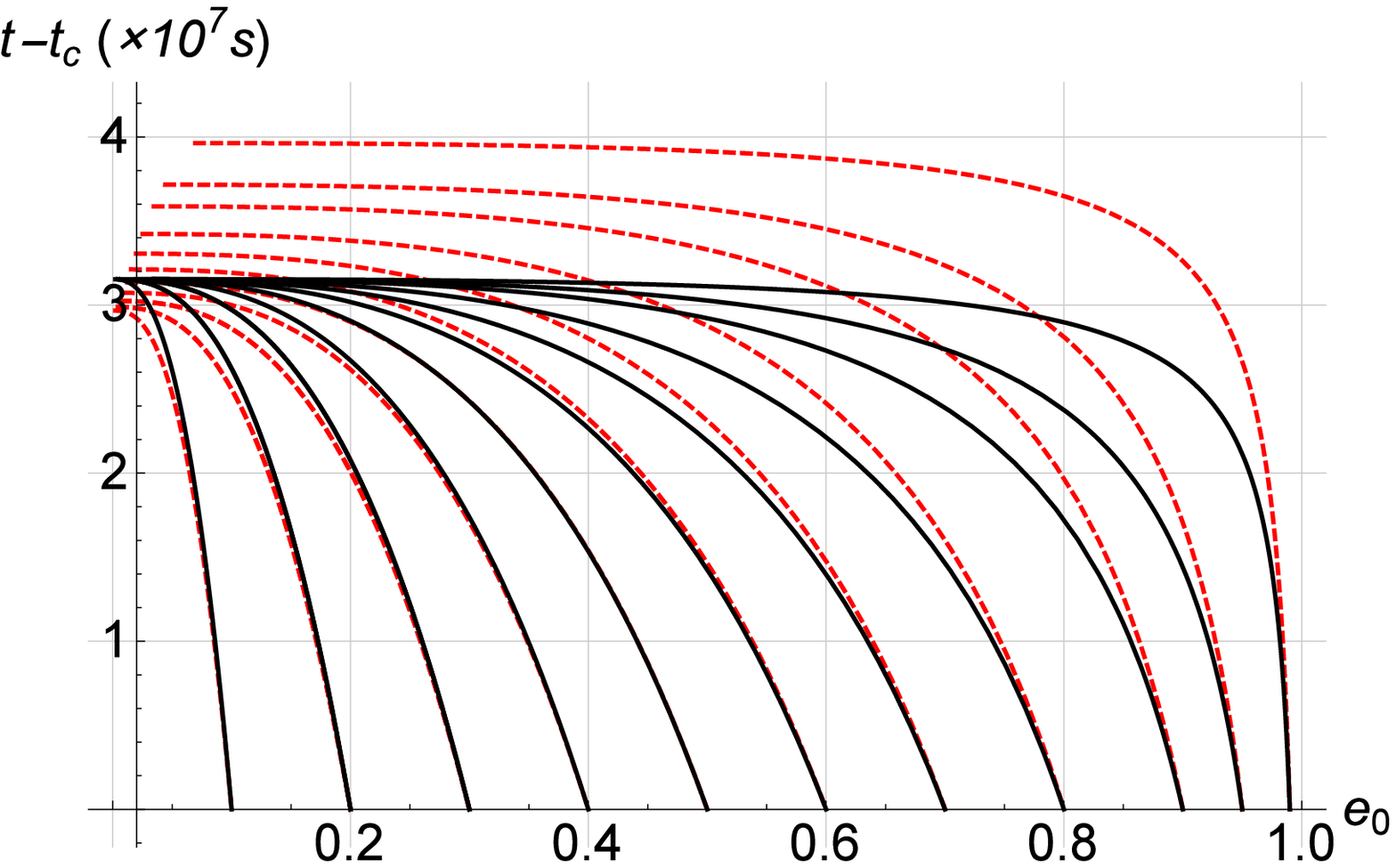}
\caption{(color online). The time function for various initial eccentricities. Newtonian and 1PN expressions are denoted by solid (black) and dashed (red) lines, respectively. The masses of the components are $m_{i}=10^{6}(1+z)M_{\odot }$ with redshift $z=1$. The inspiral time is set to 1 year before LSO calculated in the Newtonian order.}
\label{fig3}
\end{figure}

\begin{figure}[h]
\includegraphics[height=4cm]{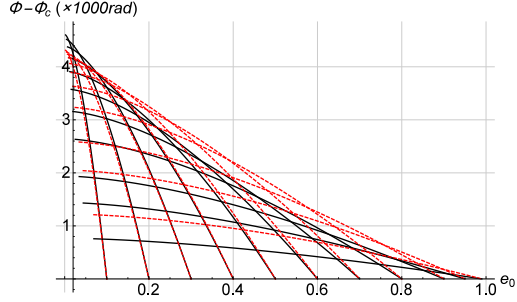}
\caption{(color online). The phase functions for various initial eccentricities.
Newtonian and 1PN expressions are denoted by solid (black) and dashed (red) lines, respectively.
The masses of the components are $m_{i}=10^{6}(1+z)M_{\odot }$ with redshift $z=1$. The
inspiral time is set to 1 year before LSO calculated in the Newtonian order.}
\label{fig4}
\end{figure}

\section{Summary}

In our work we have investigated the orbital evolution and emitted radiation of binary systems on eccentric orbits up to 1PN order. Both the time and frequency domain waveforms are presented in a simple form with use the generalized true anomaly parameterization. To express the time dependence of the waveforms the Hansen coefficients were generalized to 1PN accuracy. Moreover, the radiation reaction problem and the evolution of the time and phase functions are given to 1PN accuracy.

\acknowledgments This work was supported by the Hungarian Scientific
Research Fund (OTKA) grant No. K101709. B.M. was supported by the
Postdoctoral Fellowship Programme, and M.V. by the J\'{a}nos Bolyai
Research Scholarship, of the Hungarian Academy of Sciences. Partial
support comes from "NewCompStar", COST Action MP1304.

\appendix

\section{Tensor spherical harmonics}

Following the notation of \cite{Mathews} the traceless, symmetric
and unit basis tensors can be written as
\begin{eqnarray}
t^{\pm 2} &=&\frac{1}{2}\left( e_{x}\otimes e_{x}-e_{y}\otimes e_{y}\right)
\pm \frac{i}{2}\left( e_{x}\otimes e_{y}+e_{y}\otimes e_{x}\right) ,  \notag
\\
t^{\pm 1} &=&\mp \frac{1}{2}\left( e_{x}\otimes e_{z}+e_{z}\otimes
e_{x}\right) -\frac{i}{2}\left( e_{y}\otimes e_{z}+e_{z}\otimes e_{y}\right)
,  \notag \\
t^{0} &=&\frac{1}{\sqrt{6}}\left( -e_{x}\otimes e_{x}-e_{y}\otimes
e_{y}+2e_{z}\otimes e_{z}\right) .
\end{eqnarray}%
The scalar harmonic tensors on this basis are given by%
\begin{equation}
T^{2l^{\prime },lm}=\underset{m^{\prime }=-l^{\prime }}{\overset{l^{\prime }}%
{\sum }}\underset{m^{\prime \prime }=-2}{\overset{2}{\sum }}(l^{\prime
}2m^{\prime }m^{\prime \prime },lm)Y^{l^{\prime }m^{\prime }}t^{m^{\prime
\prime }},  \label{scalar}
\end{equation}%
where $(l^{\prime }2m^{\prime }m^{\prime \prime },lm)$ denotes the
\textit{Clebsch-Gordan coefficients} and $Y^{lm}$ is the
conventional spherical harmonic. Then the \textit{electric} and
\textit{magnetic} tensor harmonics
can be expresssed as%
\begin{eqnarray}
T^{E2,lm} &=&\sqrt{\frac{l(l+1)}{2(2l+1)(2l+3)}}T^{2\text{ }l+2,lm}+\sqrt{%
\frac{3(l-1)(l+2)}{(2l+1)(2l+3)}}T^{2\text{ }l,lm}  \notag \\
&&+\sqrt{\frac{(l+1)(l+2)}{2(2l-1)(2l+1)}}T^{2\text{ }l-2,lm}, \\
T^{B2,lm} &=&-i\sqrt{\frac{l-1}{2l+1}}T^{2\text{ }l+1,lm}-i\sqrt{\frac{l+2}{%
2l+1}}T^{2\text{ }l-1,lm}.
\end{eqnarray}
As an example we consider the tensor harmonics $T^{E2,22}$ and $T^{B2,22}$
appearing in the Newtonian waveform. Using the relationship between the
Descartes and spherical polar coordinates,%
\begin{eqnarray}
e_{x} &=&e_{r}\sin \theta \cos \varphi +e_{\theta }\cos \theta \cos \varphi
-e_{\varphi }\sin \varphi ,  \notag \\
e_{y} &=&e_{r}\sin \theta \sin \varphi +e_{\theta }\cos \theta \sin \varphi
+e_{\varphi }\cos \varphi ,  \notag \\
e_{z} &=&e_{r}\cos \theta -e_{\theta }\sin \theta ,
\end{eqnarray}
the tensor harmonics have the form%
\begin{eqnarray}
T^{E2,22} &=&\frac{1}{8}\sqrt{\frac{5}{2\pi }}[\left( 1+\cos ^{2}\theta
\right) h_{+}+2i\cos \theta h_{\times }]e^{2i\varphi }, \\
T^{B2,22} &=&-\frac{1}{16}\sqrt{\frac{5}{2\pi }}[\left( 3+\cos 2\theta
\right) h_{\times }-4i\cos \theta h_{+}]e^{2i\varphi },
\end{eqnarray}
where $h_{+}=e_{\theta }\otimes e_{\theta }-e_{\varphi }\otimes e_{\varphi }$
and $h_{\times }=e_{\theta }\otimes e_{\varphi }+e_{\varphi }\otimes
e_{\theta }$ are the two independent polarizations. The tensor spherical
harmonics up to 2PN are given in \cite{Tessmer2}.

\section{Orbital parameters of the 1PN dynamics}

The orbital parameters up to 1PN order are given by \cite{DD}
\begin{align}
n& =\frac{\left( -A\right) ^{3/2}}{B}\mathbf{,}  \notag \\
a_{R}& =-\frac{B}{A}+\frac{D}{2C_{0}}\mathbf{,}  \notag \\
e_{t}& =\left[ 1-\frac{A}{B^{2}}\left( C-\frac{BD}{C_{0}}\right) \right]
^{1/2}\mathbf{,}  \notag \\
e_{R}& =\left( 1-\frac{AD}{2BC_{0}}\right) e_{t}\mathbf{,}  \notag \\
e_{\theta }& =\left( 1+\frac{AD}{BC_{0}}-\frac{AI}{BH}\right) e_{t}\mathbf{,}
\notag \\
k& =\frac{3Gm}{a_{r}(1-e_{R}^{2})},
\end{align}
with the quantities $A-K$ are
\begin{align}
A& =\frac{2E}{\mu }\left( 1+\frac{3}{2}\left( 3\eta -1\right) \frac{E}{\mu
c^{2}}\right) \mathbf{,}  \notag \\
B& =GM\left( 1+\left( 7\eta -6\right) \frac{E}{\mu c^{2}}\right) \mathbf{,}
\notag \\
C& =-\frac{J^{2}}{\mu ^{2}}\left( 1+2\left( 3\eta -1\right) \frac{E}{\mu
c^{2}}\right) +\left( 5\eta -10\right) \frac{G^{2}M^{2}}{c^{2}}\mathbf{,}
\notag \\
D& =\left( -3\eta +8\right) \frac{GMJ^{2}}{\mu ^{2}c^{2}}\mathbf{,}  \notag
\\
H& =\frac{J}{\mu }\left( 1+\left( 3\eta -1\right) \frac{E}{\mu c^{2}}\right)
\mathbf{,}  \notag \\
I& =\left( 2\eta -4\right) \frac{GMJ}{\mu c^{2}}\mathbf{,}  \notag \\
K& =\frac{J}{\mu \left( \frac{J^{2}}{\mu ^{2}}-\frac{6G^{2}M^{2}}{c^{2}}%
\right) ^{1/2}}.
\end{align}

\section{Hansen coefficients}

The Hansen coefficients are important functions of the celestial mechanics
which are known for more than 100 years. The expansion of
Hansen-coefficients is
\begin{equation}
\left( \frac{r}{a}\right) ^{n}\exp (im\phi )=\overset{\infty }{\underset{%
k=-\infty }{\sum }}X_{k}^{n,m}\exp (ik\mathcal{M}),
\end{equation}%
where $r$ is the relative distance, $a$ is the semimajor axis, $\phi $ is
the true anomaly and $\mathcal{M}$ is the mean anomaly using by the standard
notations of celestial mechanics. The coefficients $X_{k}^{n,m}$ are called
the \textit{Hansen-coefficients}. Here the constants $n$ and $m$ are
integers. The Fourier series representation of the Hansen coefficients is
\begin{equation}
X_{k}^{n,m}=\frac{1}{2\pi }\underset{-\pi }{\overset{\pi }{\int }}\left(
\frac{r}{a}\right) ^{n}\exp (im\phi -ik\mathcal{M})d\mathcal{M}.  \label{int}
\end{equation}
The integrand can be transformation to other argument, specifically
the eccentric and true anomalies with the use of the leading order
Kepler-equation,
\begin{eqnarray}
\frac{d\mathcal{M}}{du} &=&1-e\cos u,  \label{dM} \\
\frac{d\mathcal{M}}{dv} &=&\frac{\left( 1-e^{2}\right) ^{3/2}}{\left(
1+e\cos \phi \right) ^{2}}.
\end{eqnarray}
We introduce the complex variables $x=\exp i\phi $ and $y=\exp iu$
($z=\exp i\mathcal{M}$ for the contour integral), then the
relationship between the eccentric and true anomalies (\cite{Erdi})
with $x$ and $y$ complex variables is
\begin{equation}
\frac{x-1}{x+1}=\frac{1+\beta }{1-\beta }\frac{y-1}{y+1},
\end{equation}
and one gets for the variable $x$
\begin{equation}
\exp i\phi =y(1-\beta y^{-1})(1-\beta y)^{-1},
\end{equation}
and the mean anomaly $\mathcal{M}$
\begin{eqnarray}
\frac{d\mathcal{M}}{du} &=&\frac{r}{a}=\left( 1+\beta ^{2}\right)
^{-1}(1-\beta y)(1-\beta y^{-1}), \\
\exp i\mathcal{M} &=&y\exp \left[ -\frac{e}{2}\left( y-y^{-1}\right) \right]
.
\end{eqnarray}
The integrand with eccentric anomaly $u$ is given as
\begin{eqnarray}
X_{k}^{n,m} &=&\frac{\left( 1+\beta ^{2}\right) ^{-n-1}}{2\pi }\underset{%
-\pi }{\overset{\pi }{\int }}y^{m-k}(1-\beta y^{-1})^{n+m+1}  \notag \\
&&\times (1-\beta y)^{n-m+1}\exp \left[ \frac{ke}{2}\left( y-y^{-1}\right) %
\right] du.  \label{intx}
\end{eqnarray}
The integral can be extended to infinity as a series of the Bessel
functions
\begin{equation}
X_{k}^{n,m}=(1+\beta ^{2})^{-n-1}\overset{\infty }{\underset{p=-\infty }{%
\sum }}E_{k-p}^{n,m}J_{p}(ke).  \label{aksenov}
\end{equation}
The coefficients $E_{l}^{n,m}$ for $l\geqq m$ and $E_{-l}^{n,-m}$ for $l<m$
can be expressed by the hypergeometric function $F(a,b;c;d)$ as
\begin{eqnarray}
E_{l}^{n,m} &=&\left( -\beta \right) ^{l-m}\binom{n-m+1}{l-m}  \notag \\
&&\times F(l-n-1,-m-n-1;l-m+1;\beta ^{2}).
\end{eqnarray}
The first description of this formula was given by Hill \cite{Plummer}. The
other representation of the Hansen coefficients can be found in the work of
Tisserand on celestial mechanics from 1889,%
\begin{equation}
X_{k}^{n,m}=\frac{\left( -\beta \right) ^{\left\vert k-m\right\vert }}{%
(1+\beta ^{2})^{n+1}}\overset{\infty }{\underset{s=0}{\sum }}\mathcal{P}_{s}%
\mathcal{Q}_{s}\beta ^{2s},  \label{tisserand}
\end{equation}
where%
\begin{eqnarray}
\mathcal{P}_{s} &=&\left\{
\begin{array}{l}
P_{s+k-m}\qquad k\geqq m \\
P_{s}\qquad \qquad k<m%
\end{array}%
\right\} , \\
\mathcal{Q}_{s} &=&\left\{
\begin{array}{l}
Q_{s}\qquad \qquad k\geqq m \\
Q_{s+m-k}\qquad k<m%
\end{array}%
\right\} ,
\end{eqnarray}%
and
\begin{eqnarray}
P_{s} &=&\overset{s}{\underset{r=0}{\sum }}\binom{n-m+1}{s-r}\frac{1}{r!}%
\left( \frac{kr}{2\beta }\right) ^{r}, \\
Q_{s} &=&\overset{s}{\underset{r=0}{\sum }}\binom{n+m+1}{s-r}\frac{1}{r!}%
\left( -\frac{kr}{2\beta }\right) ^{r}.
\end{eqnarray}

\section{Leading and half order waveforms}

The leading order waveform with the true anomaly $\phi $ due to
Einstein quadrupole formula are (the notation of ref. \cite{mikoczi}
for azimuthal polar angle is $\gamma =\Phi $)
\begin{eqnarray}
h_{+}^{N}(\phi ) &=&\frac{1}{1-e^{2}}\sum_{m=0}^{5}\left( c_{m}^{N+}\cos
m\phi +s_{m}^{N+}\sin m\phi \right) , \\
h_{\times }^{N}(\phi ) &=&\frac{1}{1-e^{2}}\sum_{m=0}^{5}\left(
c_{m}^{N\times }\cos m\phi +s_{m}^{N\times }\sin m\phi \right) ,
\end{eqnarray}
where
\begin{align}
c_{0}^{N+}& =-\frac{e^{2}}{2}(1+3\cos 2\Phi -2\cos 2\Theta \sin ^{2}\Phi ),
& c_{0}^{N\times }& =2e^{2}\cos \Theta \sin 2\Phi ,  \notag \\
c_{1}^{N+}& =\frac{e}{2}(-1+\cos 2\Theta -5(1+\cos ^{2}\Theta )\cos 2\Phi ),
& c_{1}^{N\times }& =5e\cos \Theta \sin 2\Phi ,  \notag \\
c_{2}^{N+}& =-(3+\cos 2\Theta )\cos 2\Phi , & c_{2}^{N\times }& =4\cos
\Theta \sin 2\Phi ,  \notag \\
c_{3}^{N+}& =-\frac{e}{4}(3+\cos 2\Theta )\cos 2\Phi , & c_{3}^{N\times }&
=e\cos \Theta \sin 2\Phi ,  \notag \\
s_{1}^{N+}& =-\frac{5e}{4}(3+\cos 2\Theta )\sin 2\Phi , & s_{1}^{N\times }&
=-5\cos \Theta \cos 2\Phi ,  \notag \\
s_{2}^{N+}& =-(3+\cos 2\Theta )\sin 2\Phi , & s_{2}^{N\times }& =-4\cos
\Theta \cos 2\Phi ,  \notag \\
s_{3}^{N+}& =-\frac{e}{4}(3+\cos 2\Theta )\sin 2\Phi , & s_{3}^{N\times }&
=-e\cos \Theta \cos 2\Phi .
\end{align}

The half order waveforms (denoted by the superscript $H$) are
\begin{eqnarray}
h_{+}^{H}(\phi ) &=&\frac{16}{\left( 1-e^{2}\right) ^{3/2}}%
\sum_{m=0}^{5}\left( c_{m}^{+}\cos m\phi +s_{m}^{+}\sin m\phi \right) , \\
h_{\times }^{H}(\phi ) &=&\frac{8}{\left( 1-e^{2}\right) ^{3/2}}%
\sum_{m=0}^{5}\left( c_{m}^{\times }\cos m\phi +s_{m}^{\times }\sin m\phi
\right) ,
\end{eqnarray}
where

\begin{eqnarray}
c_{0}^{H+} &=&-e\left( 2\left( -11+2e^{2}+\left( -1+6e^{2}\right) \cos
2\Theta \right) \sin \Theta \sin \Phi -2e^{2}(5\sin \Theta +\sin 3\Theta
)\sin 3\Phi \right) ,  \notag \\
c_{1}^{H+} &=&-\frac{1}{4}\left( 2\left( -44-19e^{2}+\left(
-4+39e^{2}\right) \cos 2\Theta ]\right) \sin \Theta \sin \Phi -35e^{2}(5\sin
\Theta +\sin 3\Theta )\sin 3\Phi \right) ,  \notag \\
c_{2}^{H+} &=&e((23\sin \Theta -5\sin 3\Theta )\sin \Phi +15(5\sin \Theta
+\sin 3\Theta )\sin 3\Phi ),  \notag \\
c_{3}^{H+} &=&-\frac{1}{4}\left( e^{2}(-23\sin \Theta +5\sin 3\Theta )\sin
\Phi -2\left( 18+7e^{2}\right) (5\sin \Theta +\sin 3\Theta )\sin 3\Phi
\right) ,  \notag \\
c_{4}^{H+} &=&5e(5\sin \Theta +\sin 3\Theta )\sin 3\Phi ,  \notag \\
c_{5}^{H+} &=&\frac{3}{4}e^{2}(5\sin \Theta +\sin 3\Theta )\sin 3\Phi ,
\notag \\
s_{1}^{H+} &=&-\frac{1}{2}\cos \Phi \left( \left( 42-72e^{2}+175e^{2}\cos
2\Phi \right) \sin \Theta +\left( 2-32e^{2}+35e^{2}\cos 2\Phi \right) \sin
3\Theta \right) ,  \notag \\
s_{2}^{H+} &=&-e((23\sin \Theta -5\sin 3\Theta )\cos \Phi +15(5\sin \Theta
+\sin 3\Theta )\cos 3\Phi ),  \notag \\
s_{3}^{H+} &=&-\frac{1}{4}\left( e^{2}(23\sin \Theta -5\sin 3\Theta )\cos
\Phi +2\left( 18+7e^{2}\right) (5\sin \Theta +\sin 3\Theta )\cos 3\Phi
\right) ,  \notag \\
s_{4}^{H+} &=&-5e(5\sin \Theta +\sin 3\Theta )\cos 3\Phi ,  \notag \\
s_{5}^{H+} &=&-\frac{3}{4}e^{2}(5\sin \Theta +\sin 3\Theta )\cos 3\Phi .
\end{eqnarray}

\section{1PN waveform}

\subsection{The coefficients proportional to $\cos m\protect\phi $}

There are coefficients proportional to $\cos m\tilde{\phi}$ and
$\sin m\tilde{\phi}$ in the 1PN waveform, Eq. (\ref{PN}).
Coefficients for $\cos 4 \tilde{\phi}$ and $\sin 4\tilde{\phi}$ are
(we have introduced the shorthand notations $c_{m\Phi }=\cos m\Phi
$, $s_{m\Phi }=\sin m\Phi $, $c_{m\Theta }=\cos m\Theta $,
$s_{m\Theta }=\sin m\Theta $, $G_{\Theta }=-5+4c_{2\Theta
}+c_{4\Theta }$ and $\lambda _{\eta }=-1+3\eta $. For example, $%
s_{3(c4)}^{+} $ mean plus polarization, proportional to $\sin 3{\phi }$ and
proportional to $\sin 4\tilde{\phi}$)
\begin{eqnarray}
c_{4(c4)}^{+} &=&\frac{1}{32}c_{4\Phi }G_{\Theta }\lambda _{\eta }\eta e^{4},
\notag \\
c_{4(s4)}^{+} &=&\frac{1}{32}s_{4\Phi }G_{\Theta }\lambda _{\eta }\eta e^{4},
\notag \\
c_{3(c4)}^{+} &=&\frac{55}{256}c_{4\Phi }G_{\Theta }\lambda _{\eta }\eta
e^{3},  \notag \\
c_{3(s4)}^{+} &=&\frac{55}{256}s_{4\Phi }G_{\Theta }\lambda _{\eta }\eta
e^{2},  \notag \\
c_{2(c4)}^{+} &=&\frac{125}{192}c_{4\Phi }G_{\Theta }\lambda _{\eta }\eta
e^{2},  \notag \\
c_{2(s4)}^{+} &=&\frac{125}{192}s_{4\Phi }G_{\Theta }\lambda _{\eta }\eta
e^{2},  \notag \\
c_{1(c4)}^{+} &=&\frac{1}{768}c_{4\Phi }G_{\Theta }\lambda _{\eta }\eta
(764+135e^{2})e,  \notag \\
c_{1(s4)}^{+} &=&\frac{1}{768}s_{4\Phi }G_{\Theta }\lambda _{\eta }\eta
(764+135e^{2})e,  \notag \\
c_{0(c4)}^{+} &=&\frac{1}{192}c_{4\Phi }G_{\Theta }\lambda _{\eta }\eta
(64+71e^{2}),  \notag \\
c_{0(s4)}^{+} &=&\frac{1}{192}s_{4\Phi }G_{\Theta }\lambda _{\eta }\eta
(64+71e^{2}).
\end{eqnarray}

The coefficients proportional to $\cos 2\tilde{\phi}$ and $\sin 2\tilde{\phi}$ are
\begin{eqnarray}
c_{3(c2)}^{+} &=&\frac{1}{64}c_{2\Phi }\left[ 291+c_{4\Theta }-3\eta
(27+c_{4\Theta })+4(19+9\eta )c_{2\Theta }\right] \eta e^{3},  \notag \\
c_{3(s2)}^{+} &=&\frac{1}{64}s_{2\Phi }\left[ 291+c_{4\Theta }-3\eta
(27+c_{4\Theta })+4(19+9\eta )c_{2\Theta }\right] \eta e^{3},  \notag \\
c_{2(c2)}^{+} &=&-\frac{1}{48}c_{2\Phi }(-983+114e^{2}+375\eta +18e^{2}\eta
+c_{4\Theta }(1+6e^{2})\lambda _{\eta }+2c_{2\Theta }(-116-81\eta
+6e^{2}(3+\eta )))e^{2}\eta ,  \notag \\
c_{2(s2)}^{+} &=&-\frac{1}{48}s_{2\Phi }(-983+114e^{2}+375\eta +18e^{2}\eta
+c_{4\Theta }(1+6e^{2})\lambda _{\eta }+2c_{2\Theta }(-116-81\eta
+6e^{2}(3+\eta )))e^{2}\eta ,  \notag \\
c_{1(c2)}^{+} &=&\frac{1}{192}c_{2\Phi }(5948+99e^{2}-3012\eta -1089\eta
e^{2}-c_{4\Theta }(4+81e^{2})\lambda _{\eta }+4c_{2\Theta }(33e^{2}(-1+\eta
)+4(76+81\eta )))e\eta ,  \notag \\
c_{1(s2)}^{+} &=&\frac{1}{192}s_{2\Phi }(5948+99e^{2}-3012\eta -1089\eta
e^{2}-c_{4\Theta }(4+81e^{2})\lambda _{\eta }+4c_{2\Theta }(33e^{2}(-1+\eta
)+4(76+81\eta )))e\eta ,  \notag \\
c_{0(c2)}^{+} &=&\frac{1}{48}c_{2\Phi }(508+e^{2}(353-465\eta )-228\eta
-c_{4\Theta }(-4+19e^{2})\lambda _{\eta }+2c_{2\Theta }(52+60\eta
+e^{2}(14+57\eta )))\eta ,  \notag \\
c_{0(s2)}^{+} &=&\frac{1}{48}s_{2\Phi }(508+e^{2}(353-465\eta )-228\eta
-c_{4\Theta }(-4+19e^{2})\lambda _{\eta }+2c_{2\Theta }(52+60\eta
+e^{2}(14+57\eta )))\eta .
\end{eqnarray}

The coefficients without $\tilde{\phi}$ dependence are
\begin{eqnarray}
c_{3}^{+} &=&-\frac{1}{256}\left[ (5-15\eta )c_{4\Theta }+11(-11+\eta
)+4(29+\eta )c_{2\Theta }\right] \eta e^{3},  \notag \\
c_{2}^{+} &=&\frac{5}{64}\left[ 37+4(-9+\eta )c_{2\Theta }-7\eta +\lambda
_{\eta }c_{4\Theta }\right] e^{2}\eta ,  \notag \\
c_{1}^{+} &=&\frac{1}{256}(1196+675e^{2}-196\eta -265e^{2}\eta +c_{4\Theta
}(-4+39e^{2})\lambda _{\eta }+4c_{2\Theta }(-300+52\eta +e^{2}(-159+37\eta
)))e\eta ,  \notag \\
c_{0}^{+} &=&\frac{1}{64}\eta (363-50e^{2}-73\eta -10e^{2}\eta +c_{4\Theta
}(-1+6e^{2})\lambda _{\eta }+c_{2\Theta }(-364-8e^{2}(-7+\eta )+76\eta )).
\end{eqnarray}

\subsection{Coefficients proportional to $\sin m\protect\phi $}

The coefficients proportional to $\cos 4\tilde{\phi}$ and $\sin 4\tilde{\phi}$ are
\begin{eqnarray}
s_{4(c4)}^{+} &=&-\frac{1}{32}s_{4\Phi }G_{\Theta }\lambda _{\eta }\eta
e^{4},  \notag \\
s_{4(s4)}^{+} &=&\frac{1}{32}c_{4\Phi }G_{\Theta }\lambda _{\eta }\eta e^{4},
\notag \\
s_{3(c4)}^{+} &=&-\frac{25}{128}s_{4\Phi }G_{\Theta }\lambda _{\eta }\eta
e^{3},  \notag \\
s_{3(s4)}^{+} &=&\frac{25}{128}c_{4\Phi }G_{\Theta }\lambda _{\eta }\eta
e^{3},  \notag \\
s_{2(c4)}^{+} &=&-\frac{15}{32}s_{4\Phi }G_{\Theta }\lambda _{\eta }\eta
e^{2},  \notag \\
s_{2(s4)}^{+} &=&-\frac{15}{32}c_{4\Phi }G_{\Theta }\lambda _{\eta }\eta
e^{2},  \notag \\
s_{1(c4)}^{+} &=&\frac{1}{128}G_{\Theta }s_{4\Phi }\lambda _{\eta
}(52+9e^{2})\eta e,  \notag \\
s_{1(s4)}^{+} &=&-\frac{1}{128}G_{\Theta }c_{4\Phi }\lambda _{\eta
}(52+9e^{2})\eta e.
\end{eqnarray}

The coefficients proportional to $\cos 2\tilde{\phi}$ and $\sin 2\tilde{\phi}$
(coefficients of $\sin m\phi $) are
\begin{eqnarray}
s_{3(c2)}^{+} &=&\frac{1}{64}s_{2\Phi }(-293-92c_{2\Theta }+c_{4\Theta
}+3\eta (29+4c_{2\Theta }-c_{4\Theta }))\eta e^{3},  \notag \\
s_{3(s2)}^{+} &=&-\frac{1}{64}c_{2\Phi }(-293-92c_{2\Theta }+c_{4\Theta
}+3\eta (29+4c_{2\Theta }-c_{4\Theta }))\eta e^{3},  \notag \\
s_{2(c2)}^{+} &=&\frac{1}{16}s_{2\Phi }(-259+81\eta +c_{4\Theta
}(-3+2e^{2})\lambda _{\eta }+2e^{2}(19+3\eta )+4c_{2\Theta }(2(-10+\eta
)+e^{2}(3+\eta )))\eta e^{2},  \notag \\
s_{2(s2)}^{+} &=&-\frac{1}{16}s_{2\Phi }(-259+81\eta +c_{4\Theta
}(-3+2e^{2})\lambda _{\eta }+2e^{2}(19+3\eta )+4c_{2\Theta }(2(-10+\eta
)+e^{2}(3+\eta )))\eta e^{2},  \notag \\
s_{1(c2)}^{+} &=&\frac{1}{64}s_{2\Phi }(-900+204\eta +5c_{4\Theta
}(-4+3e^{2})\lambda _{\eta }+11e^{2}(1+21\eta )+4c_{2\Theta }(-4(17+\eta
)+e^{2}(1+19\eta )))\eta e,  \notag \\
s_{1(s2)}^{+} &=&-\frac{1}{64}s_{2\Phi }(-900+204\eta +5c_{4\Theta
}(-4+3e^{2})\lambda _{\eta }+11e^{2}(1+21\eta )+4c_{2\Theta }(-4(17+\eta
)+e^{2}(1+19\eta )))\eta e.
\end{eqnarray}

The coefficients proportional to $\cos 4\tilde{\phi}$ and $\sin
4\tilde{\phi}$ are (we have introduced the notation $H_{\Theta
}=c_{\Theta}-c_{3\Theta }$)
\begin{eqnarray}
c_{4(c4)}^{\times } &=&\frac{1}{4}s_{4\Phi }H_{\Theta }\lambda _{\eta }\eta
e^{4},  \notag \\
s_{4(c4)}^{\times } &=&-\frac{1}{4}s_{4\Phi }H_{\Theta }\lambda _{\eta }\eta
e^{4},  \notag \\
c_{3(c4)}^{\times } &=&\frac{55}{64}s_{4\Phi }H_{\Theta }\lambda _{\eta
}\eta e^{3},  \notag \\
s_{3(c4)}^{\times } &=&-\frac{55}{64}c_{4\Phi }H_{\Theta }\lambda _{\eta
}\eta e^{3},  \notag \\
c_{2(c4)}^{\times } &=&\frac{125}{48}s_{4\Phi }H_{\Theta }\lambda _{\eta
}\eta e^{2},  \notag \\
s_{2(c4)}^{\times } &=&-\frac{125}{48}c_{4\Phi }H_{\Theta }\lambda _{\eta
}\eta e^{2},  \notag \\
c_{1(c4)}^{\times } &=&\frac{1}{182}s_{4\Phi }H_{\Theta }\lambda _{\eta
}\eta (764+135e^{2})e,  \notag \\
s_{1(c4)}^{\times } &=&-\frac{1}{182}c_{4\Phi }H_{\Theta }\lambda _{\eta
}\eta (764+135e^{2})e.
\end{eqnarray}
The coefficients proportional to $\cos 2\tilde{\phi}$ and $\sin 2\tilde{\phi}$ are
\begin{eqnarray}
c_{3(c2)}^{\times } &=&\frac{1}{32}s_{2\Phi }\left[ (3-9\eta )c_{3\Theta
}+11(-17+3\eta )c_{\Theta }\right] \eta e^{3},  \notag \\
s_{3(c2)}^{\times } &=&-\frac{1}{32}c_{2\Phi }\left[ (3-9\eta )c_{3\Theta
}+11(-17+3\eta )c_{\Theta }\right] \eta e^{3},  \notag \\
c_{2(c2)}^{\times } &=&\frac{1}{24}s_{2\Phi }\left[ (-17+6e^{2})\lambda
_{\eta }c_{3\Theta }+(-625+159\eta +6e^{2}(13+\eta ))c_{\Theta }\right] \eta
e^{2},  \notag \\
s_{2(c2)}^{\times } &=&-\frac{1}{24}c_{2\Phi }\left[ (-17+6e^{2})\lambda
_{\eta }c_{3\Theta }+(-625+159\eta +6e^{2}(13+\eta ))c_{\Theta }\right] \eta
e^{2},  \notag \\
c_{1(c2)}^{\times } &=&\frac{1}{96}s_{2\Phi }\left[ 5c_{3\Theta
}(-28+9e^{2})\lambda _{\eta }+(3(7+155\eta )e^{2}+4(-931+321\eta ))c_{\Theta
}\right] \eta e,  \notag \\
s_{1(c2)}^{\times } &=&-\frac{1}{96}c_{2\Phi }\left[ 5c_{3\Theta
}(-28+9e^{2})\lambda _{\eta }+(3(7+155\eta )e^{2}+4(-931+321\eta ))c_{\Theta
}\right] \eta e.
\end{eqnarray}

\end{document}